\documentclass[aps,prb,twocolumn,superscriptaddress,showpacs,amsmath,amssymb]{revtex4}
\usepackage{amsfonts}
\usepackage{graphicx}
\usepackage{dcolumn}
\usepackage{bm}

\newcommand{\be}{\begin{equation}}
\newcommand{\ee}{\end{equation}}
\newcommand{\bea}{\begin{eqnarray}}
\newcommand{\eea}{\end{eqnarray}}
\newcommand{\abs}[1]{\left\vert#1\right\vert}

\newcommand{\mfb}{\mathfrak{b}}
\newcommand{\mfbq}{\bar{\mathfrak{b}}}

\begin{document}

\title{Exact calculation of the magnetocaloric effect in the spin-1/2 $XXZ$ chain}

\author{Christian Trippe}
\affiliation{Theoretische Physik, Universit\"at Wuppertal,
          Gau{\ss}-Str.\ 20, 42097, Wuppertal, Germany}
\author{Andreas Honecker}
\affiliation{Institut f\"ur Theoretische Physik,
          Georg-August-Universit\"at G\"ottingen,
          37077 G\"ottingen, Germany}
\author{Andreas Kl\"umper}
\affiliation{Theoretische Physik, Universit\"at Wuppertal,
          Gau{\ss}-Str.\ 20, 42097, Wuppertal, Germany}
\author{Vadim Ohanyan}
\affiliation{Department of Theoretical Physics, Yerevan State University,
Al.\ Manoogian 1, 0025, Yerevan, Armenia}
\affiliation{Yerevan Physics Institute, Alikhanian Brothers 2, 0036, Yerevan,  Armenia}

\begin{abstract}
We calculate the entropy and cooling rate of the antiferromagnetic 
spin-1/2 $XXZ$ chain under an adiabatic demagnetization process using the 
quantum transfer-matrix technique and non-linear integral equations. The 
limiting case of the Ising chain (corresponding to infinitely large 
anisotropy) is presented for comparison. Our exact results 
for the Heisenberg chain are used as a crosscheck for the numerical 
exact diagonalization as well as Quantum Monte Carlo simulations and allow us to 
benchmark the numerical methods. Close to field-induced quantum phase 
transitions we observe a large magnetocaloric effect. Furthermore, we 
verify universal low-temperature power laws in the cooling rate and 
entropy, in particular linear scaling of entropy with temperature $T$ in 
the gapless Luttinger-liquid state and scaling as $\sqrt{T}$ at 
field-induced transitions to gapped phases.
\end{abstract}

\date{September 18, 2009; revised December 24, 2009}


\maketitle

\section{Introduction}

The magnetocaloric effect (MCE) in general addresses the change of
temperature of magnetic systems under the variation of an external magnetic
field. The MCE has been known since the end of the 19th
century,\cite{Warburg} and it has attracted renewed interest
recently because of potential room-temperature cooling applications
(see Refs.\ \onlinecite{rev1,rev2} for recent reviews).
On the other hand, adiabatic demagnetization is a standard
low-temperature method:
demagnetization of paramagnetic salts was the first method to reach temperatures
below $1$K\cite{MacDougall} whereas demagnetization of nuclear
spins has reached record low temperatures down to 100pK\cite{Oja,Knuuttila}
and is still the cooling method of choice in the $\mu$K-range.\cite{Strehlow}
The cooling rate at the
adiabatic demagnetization $\left( \frac{\partial T}{\partial H} \right)_S$ for
an ideal paramagnet ({\it i.e.}, a system of non-interacting magnetic
moments) is equal to $T/H$, which means linear monotonic dependence of
temperature on the magnetic field magnitude. The latter is a direct
consequence of the fact that for any paramagnetic system the entropy depends
only on the ratio $H/T$, so for any isentrope one gets
$H/T=\mbox{const}$. However, the matter could undergo crucial changes for
systems of interacting spins. For instance, in ferromagnets near the Curie
point one can observe a substantial enhancement of the effect.\cite{tishin}

As has been shown in early investigations, quantum antiferromagnets are more efficient
low-temperature magnetic coolers
than ferromagnets.\cite{Bonner1972,Bonner1962,Bonner1977}
This fact is connected with the behavior of the entropy
of antiferromagnets.
The entropy of any antiferromagnet at low temperatures displays (at least) one maximum as a function of magnetic field, which
usually, according to the third law of thermodynamics, falls to zero at $T \to
0$.\cite{Bonner1962} The Ising model is anomalous in this respect,
because of non-vanishing zero-temperature entropy at the critical magnetic
field $H_c=q J$, where $q$ is the coordination number of the lattice and $J$ is
the coupling constant.
Indeed, understanding the influence of quantum fluctuations
seems to have been an important motivation for the numerical work
Ref.~\onlinecite{Bonner1962}.
The first exact result concerning  magnetocaloric properties of
the spin-1/2, Ising-like $XXZ$ chain has been obtained in
Ref.~\onlinecite{Bonner1977}, where the isentropes in the $(H,T)$
plane have been presented. The main feature of the isentropes
of the Ising-like $XXZ$ chain is the appearance of two minima.
To the best of our knowledge, the isentropes of the
isotropic spin-1/2 Heisenberg chain have so far been investigated
only numerically\cite{zhhon} (for numerical studies of the
magnetocaloric effect in ferrimagnetic spin chains and higher-spin
Heisenberg chains see Refs.~\onlinecite{BBO07,HoWe09}).
Recent measurements\cite{Tsui} of the adiabatic cooling rate in the
spin-1/2 Heisenberg chain compound [Cu($\mu$-C$_2$O$_4$)(4-aminopyridine)$_2$(H$_2$O)]$_n$
render the magnetocaloric effect of the spin-1/2 Heisenberg chain a
topic of current interest.

More generally, the MCE is particularly large in the vicinity
of quantum critical points (QCPs). The MCE is closely related
to the generalized Gr\"{u}neisen ratios
\begin{eqnarray}
\Gamma_r=-\frac{1}{T}\frac{\left(\partial S/\partial r \right)_T}{\left(\partial S/\partial T \right)_r}\,. \label{Grun}
\end{eqnarray}
Here $r$ is the control parameter governing the quantum phase transition.
In the case of the MCE $r$ is the external magnetic field $H$.
Using basic thermodynamic relations,\cite{rev2} the generalized
Gr\"{u}neisen ratio $\Gamma_H$ can be related to the
adiabatic cooling rate $\left(\partial T/\partial H \right)_S$:\cite{Zhu,Garst}
\bea
\Gamma_H=\frac{1}{T}\left(
\frac{\partial T}{\partial H} \right)_S
= -\frac{1}{C_H}\left( \frac{\partial M}{\partial T}
\right)_H \, . \label{ncr}
\eea
Thus, the magnetic cooling rate is an important quantity for
the characterization of QCPs, {\it i.e.}, quantum phase transitions
between different magnetic structures under tuning the magnetic field at $T=0$.

In passing we mention that
an analysis of classical spin models\cite{zh03} demonstrated that
the MCE can be enhanced by geometric frustration. Indeed,
adiabatic demagnetization experiments on the frustrated spin-7/2
pyrochlore-type magnet Gd$_2$Ti$_2$O$_7$ have shown substantial drops in
temperature in the vicinity of the saturation field.\cite{sos}
Enhanced cooling performance is also theoretically predicted
in one-dimensional quantum antiferromagnets such as
the $J_1-J_2$-chain and the sawtooth chain,\cite{zhhon}
the diamond chain,\cite{DeRi06,Canova,Derzhko07,per09,can09}
as well as in two dimensions.\cite{hon06,sha07}

The one-dimensional spin-1/2 Heisenberg model is famous for its
integrability. The conventional Bethe ansatz technique allows one to obtain all
eigenvalues and eigenvectors of the corresponding Hamiltonian, though in
non-explicit form. There are several sophisticated methods to describe
thermodynamics of one-dimensional integrable models, like thermodynamic Bethe
ansatz (TBA), etc.\ (see for example Refs.~\onlinecite{Gaudin1971,tak}). For pragmatical reasons,
the most suitable technique is the quantum transfer-matrix (QTM) method
leading to only two non-linear integral equations (NLIE) for the free energy
of the Heisenberg chain, see Refs.~\onlinecite{klu98,klu00,klu} and references therein.
A numerical solution of these NLIEs has been used in Ref.~\onlinecite{klu98}
to obtain the magnetic susceptibility and specific heat of the isotropic
spin-1/2 Heisenberg chain in an external magnetic field. However, as
far as we are aware, there are no exact results for the free energy (or
equivalently the entropy) and a mixed derivative of the free energy which
corresponds to the magnetic cooling rate in the literature for
the isotropic spin-1/2 Heisenberg chain. Filling these gaps is one
of the purposes of the present paper.

In this paper we present calculations of quantities related to the MCE for the
one-dimensional spin-1/2 $XXZ$ Heisenberg chain within the QTM and NLIE
method. Particularly, isentropes for different values of the exchange
anisotropy parameter $\Delta$ are obtained. The cooling rate is computed
as a function of external
magnetic field for various fixed values of temperature and different values of
the exchange anisotropy. The limiting case of infinitely large
anisotropy which is just the Ising model is considered as well. In that case
the expressions for all relevant thermodynamic quantities can be obtained
in closed form. These calculations are supplemented by exact diagonalization
(ED) and Quantum Monte Carlo
(QMC) calculations demonstrating full agreement between results obtained from
the exact solution and numerical calculations.

\section{Isentropes and Cooling rate for the spin-1/2 $XXZ$ Heisenberg chain}

\label{sec:Heis}

We will be interested both in the entropy and the associated isentropes
as well as the temperature derivatives thereof, {\it i.e.},
the adiabatic cooling rate $\left(
\frac{\partial T}{\partial H} \right)_S$. Using standard thermodynamic
relations one can express the latter as follows:\cite{rev2}
\bea \left( \frac{\partial
  T}{\partial H} \right)_S=-\frac{T}{C_H}\left( \frac{\partial M}{\partial T}
\right)_H \, , \label{cr}
\eea
where $C_H$ is the specific heat at constant magnetic field, and $M$ is the
magnetization of the system. After normalization with a factor $1/T$,
the magnetic cooling rate can be identified with the
generalized Gr\"uneisen ratio Eq.~(\ref{ncr})
which we will use in the following.

We will be specifically interested in the spin-1/2 $XXZ$ Heisenberg chain
whose Hamiltonian is given by
\bea
\mathcal{H}&=&J\sum_{n=1}^N \left(S_n^xS_{n+1}^x+S_n^yS_{n+1}^y
+\Delta \, S_n^z S_{n+1}^z\right) \nonumber \\
&&-H\sum_{n=1}^N S_n^z \, . \label{ham}
\eea
Here $N$ is the total number of sites, $S_n^\alpha$ are spin-1/2
operators acting at site $n$, $J$ is the exchange constant,
$\Delta$ an exchange asymmetry, and $H$ an external magnetic
field. For the finite-size computations we will assume
periodic boundary conditions, {\it i.e.},
$S_{N+1}^\alpha = S_1^\alpha$. Note that the properties
of the $XXZ$ model (\ref{ham}) are symmetric under $H \to -H$
and $M = \langle S^z_n \rangle \to -M$. We will therefore concentrate
on $H \ge 0$ in the following.

\begin{figure}[t!]
\includegraphics[width=\columnwidth]{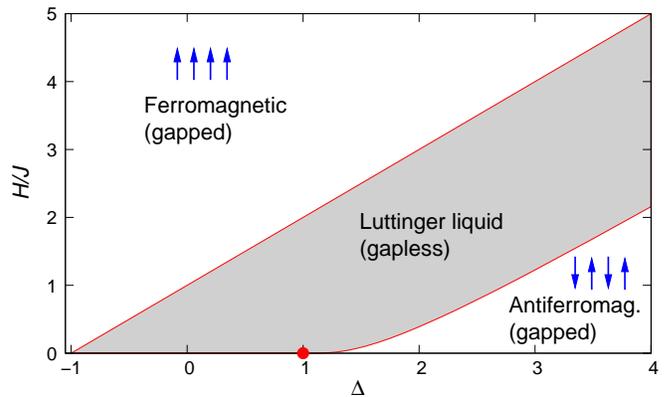}
\caption{(Color online) Zero-temperature phase diagram of the
spin-1/2 $XXZ$ chain in a magnetic field.
\label{figXXZ}}
\end{figure}

It is useful to recall the zero-temperature phase diagram
of the model Eq.~(\ref{ham}), see Fig.\ \ref{figXXZ}.\cite{JMcC72,AlMa95,CHP98}
For $\Delta > 1$ and small magnetic fields $H < H_{c1}$ there
is long-range antiferromagnetic order along the $z$-direction
in spin space.
This state exhibits a gap in the excitation spectrum
whose value at $H=0$ corresponds to
$H_{c1}$. The value of the gap has been computed exactly via the
Bethe ansatz.\cite{CG66,yang_yang_1966_3} In particular, it is exponentially small
close to the Heisenberg point at $\Delta=1$ (marked by the dot in Fig.\
\ref{figXXZ}), which is characteristic for a Kosterlitz-Thouless
transition.\cite{KT73,Kosterlitz} For $\Delta<1$, one finds
a gapless Luttinger-liquid state.\cite{Haldane}
This Luttinger-liquid state exists for $\abs{\Delta} < 1$, $H<H_{c2}$ and
$\Delta>1$, $H_{c1} < H < H_{c2}$, with $H_{c2} = J\,(1+\Delta)$.
Finally, for $H > H_{c2}$, the ground state is the ferromagnetically
polarized state along the $z$-direction which exhibits again a gap.
We anticipate that these different zero-temperature regions, in particular the
quantum phase transitions at $H_{c1}$ and $H_{c2}$ will be
reflected by the magnetocaloric properties at finite temperature.

\begin{widetext}

\subsection{Non-linear integral equations}
\label{sec:NLIE}

In this section we recall the approach to the thermodynamics
using the QTM technique and NLIEs.\cite{klu,klu98}
The equations look different for $\abs{\Delta} < 1$ and $\abs{\Delta} > 1$.

First we consider the case $\abs{\Delta} < 1$. One can represent the free energy of the system per lattice site in the following form:
\bea
f=e_0-T
\int\limits_{-\infty}^{\infty}\frac{\ln\left[(1+a(x))(1+\overline{a}(x))\right]}{4 \cosh \frac{\pi x}{2}}\,dx \label{f}
\eea
(the constant $e_0$ is irrelevant in the present context).
The auxiliary functions $a(x)$ and $\overline{a}(x)$ are found from
the following system of integral equations
\begin{subequations}
\label{uu}
\bea
\ln a(x) &=& -\frac{J\sin \gamma}{T 2 \gamma}\frac{\pi}{\cosh \frac{\pi x}{2}}+\frac{\pi H}{2 T \left(\pi -\gamma\right)}+\kappa*\ln
(1+a)(x)-\kappa^+*\ln (1+\overline{a})(x) \, , \\
\ln \overline{a}(x)&=&-\frac{J\sin \gamma}{T 2 \gamma}\frac{\pi}{\cosh \frac{\pi
x}{2}}-\frac{\pi H}{2 T
\left(\pi-\gamma\right)}+\kappa*\ln(1+\overline{a})(x)-\kappa^-*\ln (1+a)(x) \, .
\eea
\end{subequations}
Here $\gamma = \arccos \Delta$, the symbol $*$ denotes convolution
$f*g(x)=\int_{-\infty}^{\infty}f(x-y)g(y)dy$ and the function $\kappa (x)$ is
defined by
\bea
\kappa(x)=\frac{1}{2 \pi}\int\limits_{-\infty}^{\infty}\frac{\sinh\left(\frac{\pi}{\gamma}-2 \right)k}{2 \cosh k \sinh\left(\frac{\pi}{\gamma}-1 \right)k}e^{i k x}dk\, , \qquad
\kappa^\pm(x) = \kappa(x\pm 2i) \, .
 \label{ker}
\eea
These equations are valid for $0<\Delta<1$. Results for negative $\Delta$ can be obtained by changing the sign of the coupling $J$.

For the case $\Delta>1$ the free energy has the form
\bea
\label{eq:fe}
f=e_0-\frac{T}{2}c\star\ln\left[(1+\mfb)(1+\mfbq)\right](0)=e_0-\frac{T}{2\pi}\int\limits_{-\pi/2}^{\pi/2}\,c(x) \ln\left[(1+\mfb(x))(1+\mfbq(x))\right] dx
\eea
with $f\star g(x)=\frac{1}{\pi}\int_{-\pi/2}^{\pi/2}dy\,f(x-y)g(y)$ denoting a convolution
with modified integration limits and
\begin{equation}
c(x)=\sum_{k=-\infty}^\infty \frac{1}{\cosh(\eta k)} e^{i2kx} \, ,
\end{equation}
with $\Delta=\cosh \eta$.
The auxiliary functions are solutions of the following integral equations
\begin{subequations}
\label{eq:nlie_massiv}
\begin{align}	
\ln\mfb(x)=&-\frac{J\sinh(\eta)}{2T}c(x)+\frac{H}{2T}+k\star\ln\left(1+\mfb(x)\right)-k^+\star\ln\left(1+\mfbq(x)\right)\\
\ln\mfbq(x)=&-\frac{J\sinh(\eta)}{2T}c(x)-\frac{H}{2T}+k\star\ln\left(1+\mfbq(x)\right)-k^-\star\ln\left(1+\mfb(x)\right)
\end{align}
\end{subequations}
with integration kernels
\begin{equation}
k(x) = \sum_{k=-\infty}^\infty \frac{e^{-\eta|k|}}{2\cosh(\eta k)} e^{i2kx}
\, , \qquad
k^\pm(x) = k(x\pm i\eta^\mp) \, .
\end{equation}
Results for $\Delta<-1$ can again be obtained by changing the sign of the coupling $J$.

The equations for the isotropic case $\Delta=1$
can be obtained by taking the limit
$\gamma\rightarrow 0$ from the case $\abs{\Delta}<1$ in \eqref{f} and \eqref{uu}
or from the case $\abs{\Delta}>1$ by changing the spectral parameter $x=\tilde{x}\eta$
and taking the limit $\eta\rightarrow 0$ in \eqref{eq:fe} and \eqref{eq:nlie_massiv}.

Having all these exact expressions one can obtain any thermodynamic
quantity of interest by iteration of the NLIE \eqref{uu} (or \eqref{eq:nlie_massiv}) and numerical
integration of the expression for the free energy \eqref{f} (or \eqref{eq:fe}).
Derivatives of the free energy with respect to $T$ and $H$ can also be
calculated. One can avoid numerical differentiation by solving the associated integral equations for the differentiated auxiliary functions, e.g. $\partial_H \ln a(x)$. Note that derivatives of $\ln a(x)$ and $\ln \overline{a}(x)$ are treated as independent functions in these equations.
As an example we will give the equations for the calculation of the magnetization per spin $M$ and the derivative of $M$ with respect to the temperature in the regime $\abs{\Delta}<1$
\bea
\label{eq:magnetization}
M=T\int\limits_{-\infty}^{\infty}\frac{1}{4 \cosh \frac{\pi x}{2}} \left(\frac{a(x)}{1+a(x)}\left[\partial_H \ln a(x)\right]+ \frac{\overline{a}(x)}{1+\overline{a}(x)} \left[\partial_H \ln \overline{a}(x)\right]\right ) dx\,.
\eea
The derivatives $\partial_H \ln a(x)$ and $\partial_H \ln \overline{a}(x)$ satisfy {\em linear} integral equations in which the auxiliary functions $a(x)$ and $\overline{a}(x)$ enter as external functions
\begin{subequations}
\bea
\partial_H \ln a(x) &=& \frac{\pi }{2 T \left(\pi -\gamma\right)}+\kappa*\frac{a}{1+a}\left[\partial_H \ln a\right](x)-\kappa*\frac{\overline{a}}{1+\overline{a}}\left[\partial_H \ln \overline{a}\right](x+2 i) \, , \\
\partial_H \ln \overline{a}(x)&=&-\frac{\pi}{2 T \left(\pi
-\gamma\right)}+\kappa*\frac{\overline{a}}{1+\overline{a}}\left[\partial_H \ln \overline{a}\right](x)-\kappa*\frac{a}{1+a}\left[\partial_H \ln a\right](x-2 i) \, .
\eea
\end{subequations}
To obtain the cooling rate $\Gamma_H$ we have to determine $\partial M /\partial
T$. In order to achieve this we have to differentiate \eqref{eq:magnetization}
with respect to $T$.
However it has turned out that in the framework of NLIEs the resulting equations in general behave numerically better if the derivatives are taken with respect
to the inverse temperature $\beta=1/T$ (we set $k_B = 1$).
\begin{align}
\nonumber
\left(\frac{\partial M}{\partial T}\right)_H&=-\beta^2\left(\frac{\partial M}{\partial \beta}\right)_H=\\\nonumber
&\int\limits_{-\infty}^{\infty}\frac{1}{4 \cosh \frac{\pi x}{2}} \left(\frac{a(x)}{1+a(x)}\left[\partial_H \ln a(x)\right]+ \frac{\overline{a}(x)}{1+\overline{a}(x)} \left[\partial_H \ln \overline{a}(x)\right]\right ) dx\\\nonumber
&-\beta\int\limits_{-\infty}^{\infty}\frac{1}{4 \cosh \frac{\pi x}{2}} \left[\left(\frac{a(x)}{1+a(x)}-\left(\frac{a(x)}{1+a(x)}\right)^2\right)\left[\partial_\beta \ln a(x)\right] \left[ \partial_H \ln a(x)\right] +\frac{a(x)}{1+a(x)}\left[ \partial_\beta\partial_H \ln a(x)\right] \right.\\
&\quad\quad\quad\left.+\left(\frac{\overline{a}(x)}{1+\overline{a}(x)}-\left(\frac{\overline{a}(x)}{1+\overline{a}(x)}\right)^2\right)\left[ \partial_\beta \ln \overline{a}(x)\right] \left[\partial_H \ln \overline{a}(x)\right] + \frac{\overline{a}(x)}{1+\overline{a}(x)}\left[ \partial_\beta \partial_H \ln \overline{a}(x)\right] \right] dx\,.
\end{align}
Here four new functions $\partial_\beta \ln a(x)$, $\partial_\beta\partial_H \ln a(x)$ and their $\overline{a}$ counterparts occur. The corresponding linear integral equations read
\begin{subequations}
\label{eq:dlnbeta}
\bea
\partial_\beta \ln a(x) &=&-\frac{J\sin \gamma}{2 \gamma}\frac{\pi}{\cosh \frac{\pi x}{2}}+ \frac{\pi H}{2 \left(\pi -\gamma\right)}+\kappa*\frac{a}{1+a}\left[\partial_\beta \ln a\right](x)-\kappa^+*\frac{\overline{a}}{1+\overline{a}}\left[\partial_\beta \ln \overline{a}\right](x) \, , \\
\partial_\beta \ln \overline{a}(x)&=&-\frac{J\sin \gamma}{2 \gamma}\frac{\pi}{\cosh \frac{\pi x}{2}}-\frac{\pi H}{2 \left(\pi
-\gamma\right)}+\kappa*\frac{\overline{a}}{1+\overline{a}}\left[\partial_\beta \ln \overline{a}\right](x)-\kappa^-*\frac{a}{1+a}\left[\partial_\beta \ln a\right](x) \, ,
\eea
\end{subequations}
and
\begin{subequations}
\label{eq:mixedaux}
\begin{align}
\nonumber
\partial_\beta \partial_H \ln a(x) =& \frac{\pi }{2 \left(\pi -\gamma\right)}+\kappa*
\left[\left(\frac{a}{1+a}-\left(\frac{a}{1+a}\right)^2\right)\left[\partial_\beta \ln a\right] \left[ \partial_H \ln a\right] +\frac{a}{1+a}\left[ \partial_\beta\partial_H \ln a\right] \right](x)\\
&\quad -\kappa^+*
\left[\left(\frac{\overline{a}}{1+\overline{a}}-\left(\frac{\overline{a}}{1+\overline{a}}\right)^2\right)\left[ \partial_\beta \ln \overline{a}\right] \left[\partial_H \ln \overline{a}\right] + \frac{\overline{a}}{1+\overline{a}}\left[ \partial_\beta \partial_H \ln \overline{a}\right] \right](x) \, , \\ \nonumber
\partial_\beta \partial_H \ln \overline{a}(x)=&-\frac{\pi}{2 \left(\pi
-\gamma\right)}+\kappa*
\left[\left(\frac{\overline{a}}{1+\overline{a}}-\left(\frac{\overline{a}}{1+\overline{a}}\right)^2\right)\left[ \partial_\beta \ln \overline{a}\right] \left[\partial_H \ln \overline{a}\right] + \frac{\overline{a}}{1+\overline{a}}\left[ \partial_\beta \partial_H \ln \overline{a}\right] \right](x) \\
&\quad -\kappa^-*
\left[\left(\frac{a}{1+a}-\left(\frac{a}{1+a}\right)^2\right)\left[\partial_\beta \ln a\right] \left[ \partial_H \ln a\right] +\frac{a}{1+a}\left[ \partial_\beta\partial_H \ln a\right] \right]
(x) \, .
\end{align}
\end{subequations}
Note that the functions $\partial_\beta \ln a(x)$ and $\partial_\beta \ln \overline{a}(x)$ already allow the calculation of the entropy per spin
\bea
\label{eq:entropy}
S=\int\limits_{-\infty}^{\infty}\frac{\ln\left[(1+a(x))(1+\overline{a}(x))\right]}{4 \cosh \frac{\pi x}{2}}\,dx - \frac{1}{T}\int\limits_{-\infty}^{\infty}\frac{1}{4 \cosh \frac{\pi x}{2}} \left(\frac{a(x)}{1+a(x)}\left[\partial_\beta \ln a(x)\right]+ \frac{\overline{a}(x)}{1+\overline{a}(x)}\left[\partial_\beta \ln \overline{a}(x)\right]\right ) dx\, .
\eea
For the calculation of the cooling rate $\Gamma_H$
the specific heat is needed in addition to the mixed derivative $\partial M/\partial T$.
It can be obtained by differentiating \eqref{eq:entropy} with respect to $\beta$ and dividing by $-T$. In the resulting equation another pair of functions $\partial_\beta^2 \ln a(x)$, $\partial_\beta^2 \ln \overline{a}(x)$ occurs, where the corresponding integral equations are derived by differentiating \eqref{eq:dlnbeta} again with respect to $\beta$ in close analogy to \eqref{eq:mixedaux}.

\end{widetext}

\subsection{Comparison with numerical results}

\begin{figure}[tb!]
\includegraphics[width=\columnwidth]{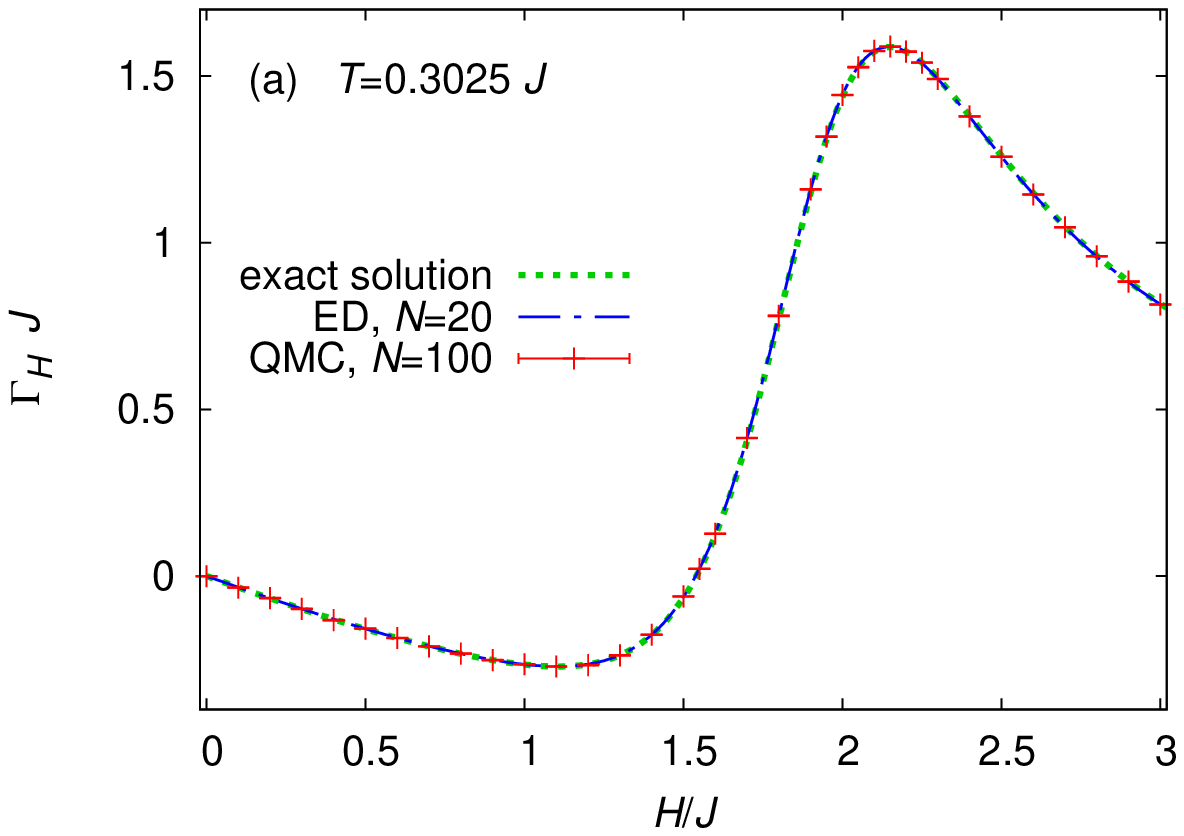}
\includegraphics[width=\columnwidth]{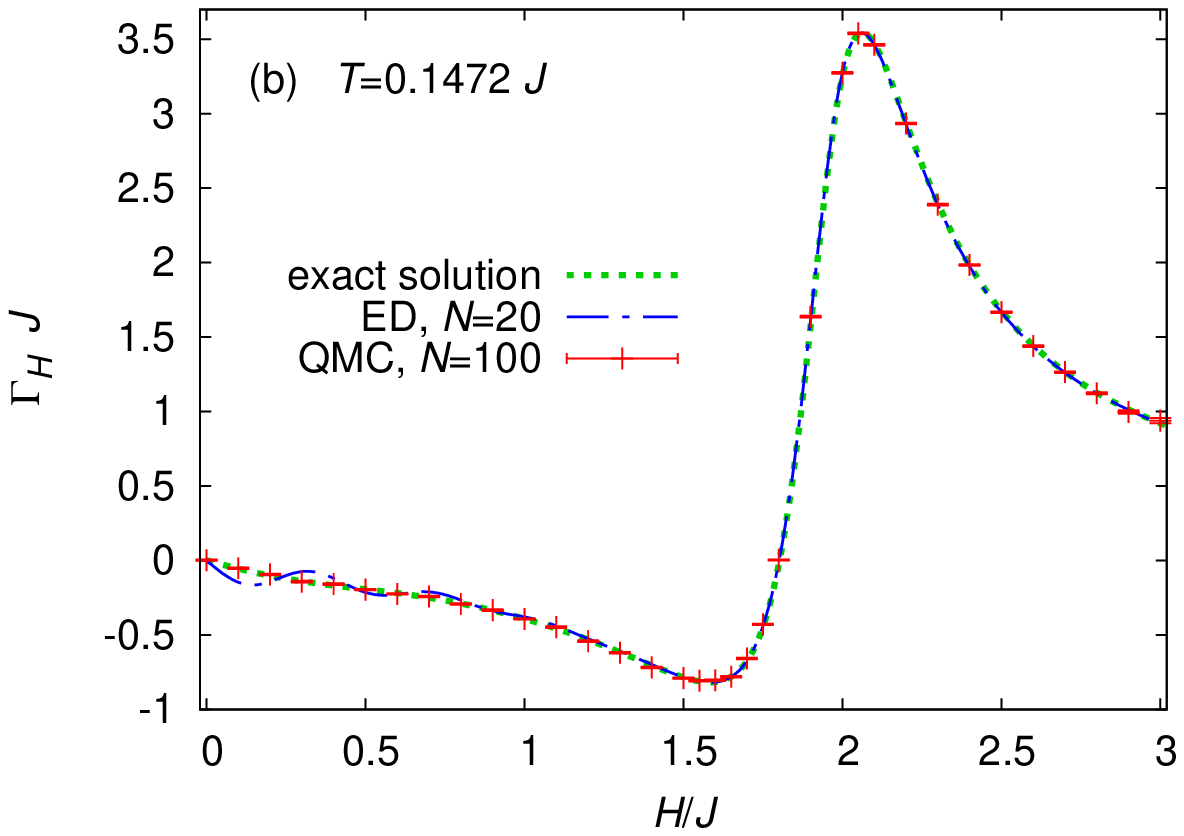}
\includegraphics[width=\columnwidth]{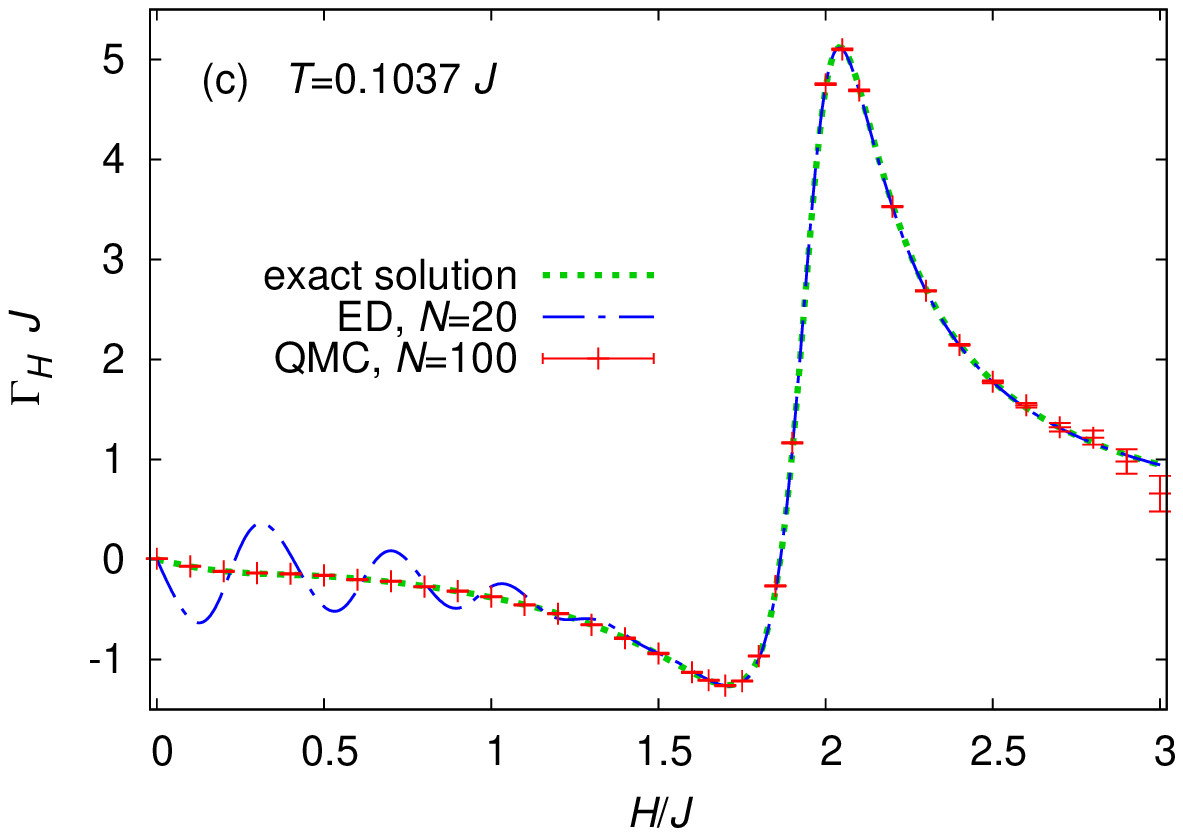}
\caption{(Color online) Cooling rate $\Gamma_H$ for the
Heisenberg chain ($\Delta =1$). Shown are
exact diagonalization (ED) for $N=20$ sites, Quantum Monte Carlo
(QMC) for $N=100$ sites and the exact solution for the infinite system.
The different panels are for $T/J=0.3025$ (a), $0.1472$ (b), and
$0.1037$ (c).
\label{fig3}}
\end{figure}

First, we present a comparison with numerical methods, namely exact diagonalization (ED) and Quantum Monte Carlo (QMC).
On the one
hand, this comparison will serve as a cross-check of our results.
On the other hand,
we can use the exact results to assess the performance of the numerical
methods.

The quantities appearing on the r.h.s.\ of Eq.~(\ref{ncr})
can be expressed as follows:
\begin{eqnarray}
C_H &=& \frac{\beta^2}{N} \, \left(
 \left\langle \mathcal{H}^2 \right\rangle - \left\langle \mathcal{H} \right\rangle^2
 \right) \, , \label{eq:CHcor} \\
\left( \frac{\partial M}{\partial T} \right)_H
    &=& \frac{\beta^2}{N} \, \left(
 \left\langle M\,\mathcal{H} \right\rangle - \left\langle M \right\rangle\left\langle \mathcal{H} \right\rangle
 \right) \, , \label{eq:dMdTcor}
\end{eqnarray}
where $\langle \cdot \rangle$ is the expectation value at a fixed
temperature $T$ and magnetic field $H$.
Here we have chosen a normalization per spin which drops out when
taking the ratio in Eq.~(\ref{ncr}). Eq.~(\ref{eq:CHcor}) is well known
and Eq.~(\ref{eq:dMdTcor}) is valid for any Hamiltonian conserving magnetization,
{\it i.e.}, $\left[M,\mathcal{H}\right] = 0$.

One can write down spectral representations for the correlation functions in
Eqs.~(\ref{eq:CHcor}) and (\ref{eq:dMdTcor}) which can be evaluated by
ED. These correlation functions can also be
evaluated with QMC.
The QMC simulations to be reported below have been
carried out with the ALPS\cite{alps1,alps2}
directed loop application\cite{alps-sse}
in the stochastic series expansion framework.\cite{Sandvik}
The specific heat Eq.~(\ref{eq:CHcor}) is measured using an improved estimator
which involves the fluctuations of the expansion order.\cite{SSS}
The correlation function Eq.~(\ref{eq:dMdTcor}) can be measured
in a similar manner.
Note that it is crucial to choose an appropriate
pseudo random-number generator in order to
obtain correct results. We have used the ``Mersenne Twister''.\cite{MTrng}
In our QMC simulations we have performed $4\cdot10^5$ thermalization steps
and then collected data during a number of sweeps ranging between
$2\cdot10^7$ and $3.3\cdot10^{10}$.

Fig.~\ref{fig3} shows a comparison of $\Gamma_H$
between the exact solution for $N=\infty$, ED for $N=20$,
and QMC for $N=100$ at
three temperatures which have been chosen to
correspond to the experiments of Ref.~\onlinecite{Tsui}.
The ED curves exhibit some wiggles at small magnetic fields and temperatures
which reflect the fact that the system contains only $N=20$ sites. In this
regime, the QMC results for $N=100$ are indistinguishable from the
exact solution for $N=\infty$ on the scale of Fig.~\ref{fig3}. On the
other hand, the QMC results are subject to big error bars at {\em high}
fields and low temperatures despite the fact that we have invested
substantial amounts of CPU time into these data points.
To some extent, this is related to performance
problems of the algorithm in a magnetic field.\cite{Sandvik} However, the
main reason is that $\Gamma_H$ is given by the ratio of two
quantities (see Eq.~(\ref{ncr})) which are both exponentially small in $T$
for $H > H_{c2} = 2\,J$. Indeed, even with a large number of sweeps, it
is difficult to determine the ratio of two very small quantities
accurately by QMC. Conversely, the existence of a gap improves finite-size
convergence such that ED works particularly well
in the high-field regime.

The overall good agreement between all three methods serves as a consistency
check for each of them. Furthermore, we see that we can get extremely accurate
numerical results by combining QMC for $H \lesssim H_{c2}$
and ED for $H \gtrsim H_{c2}$ ($H_{c2} =2\,J$ in the present case).

\subsection{Effects of anisotropy $\Delta$}

\begin{figure}[tb!]
\includegraphics[width=\columnwidth]{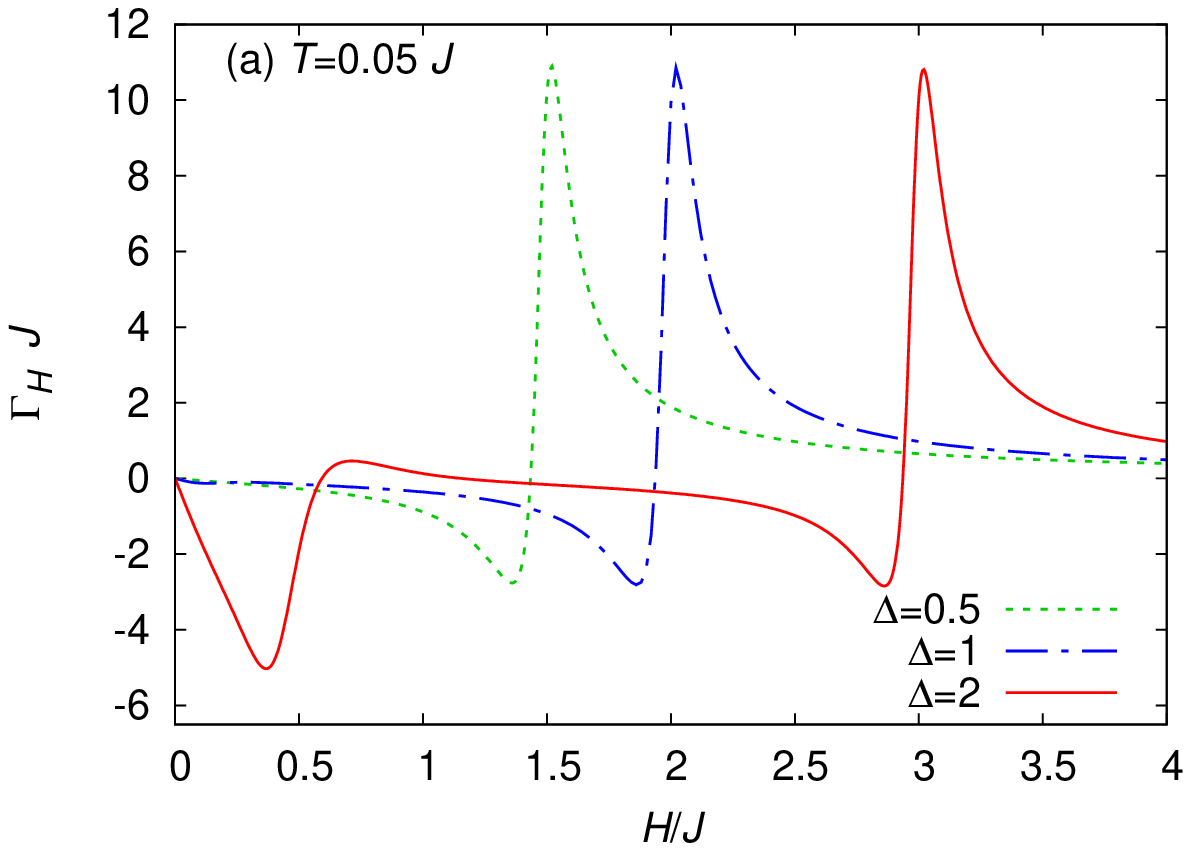}
\includegraphics[width=\columnwidth]{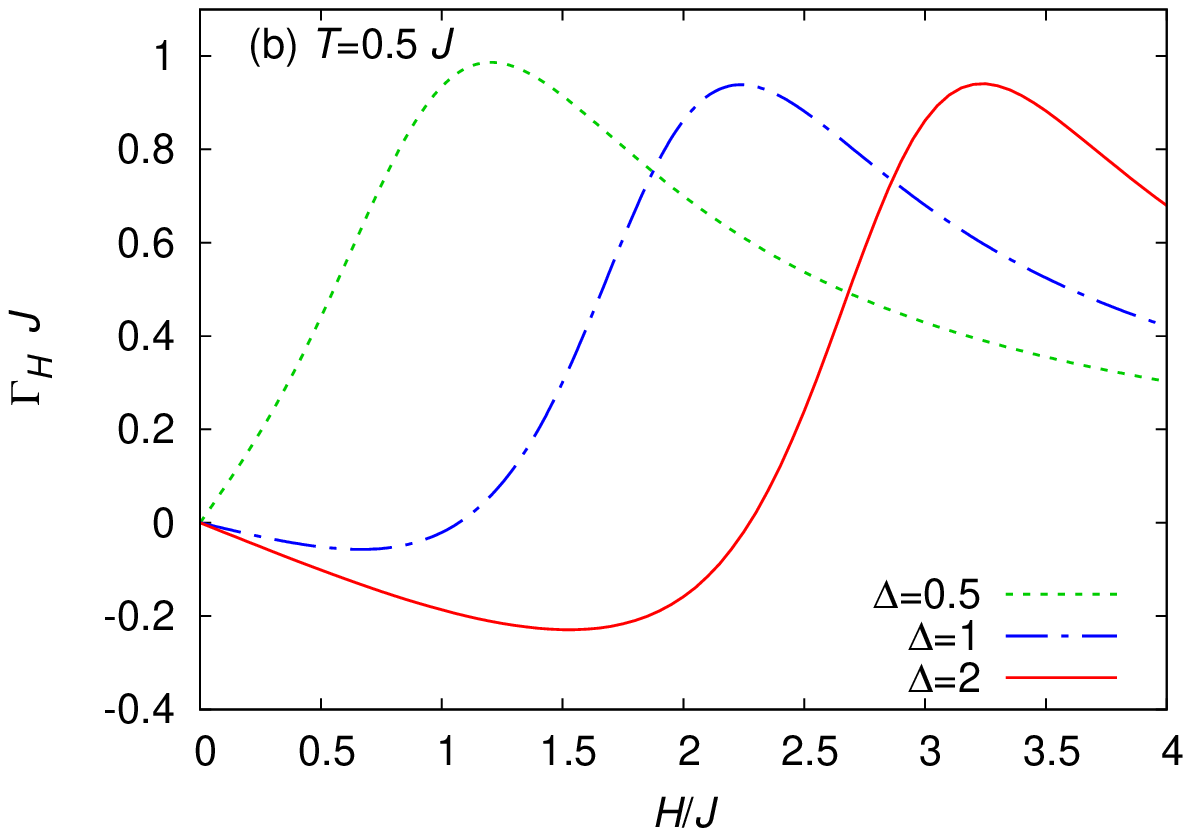}
\includegraphics[width=\columnwidth]{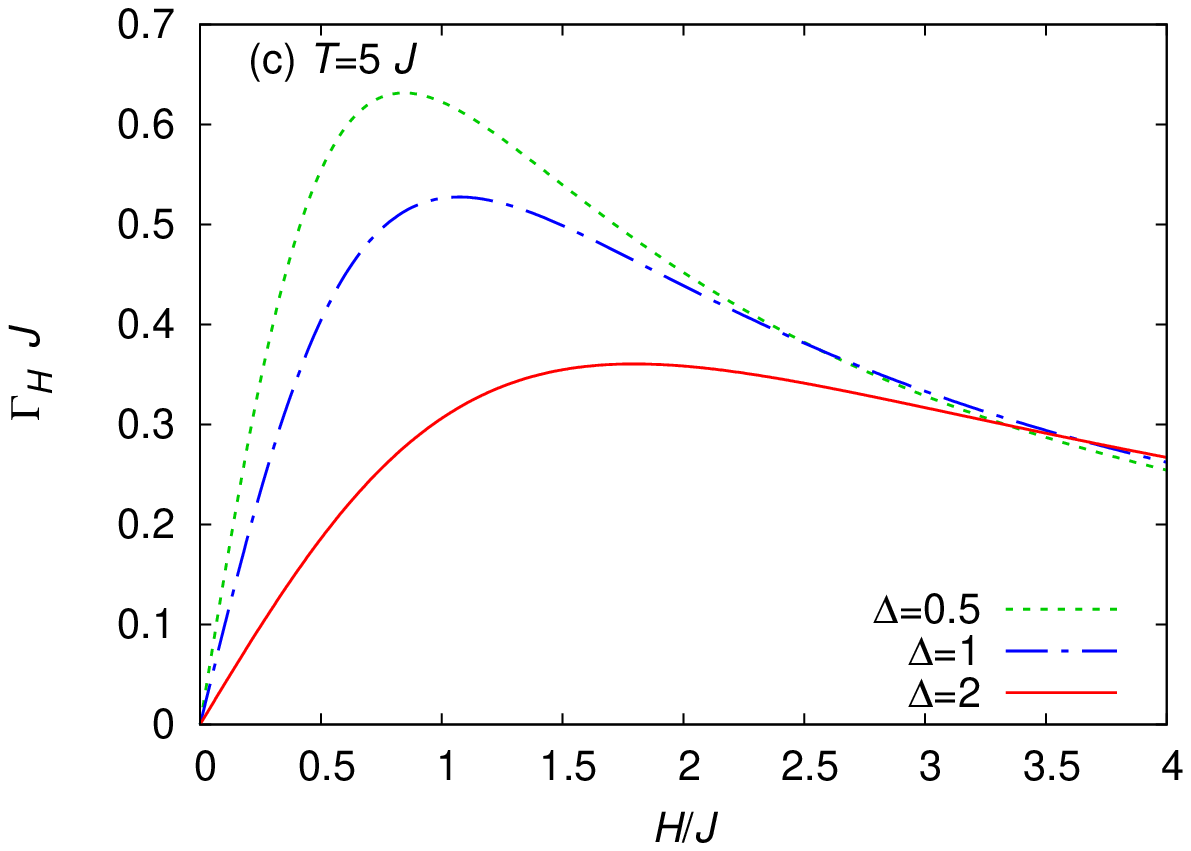}
\caption{(Color online) Cooling rate $\Gamma_H$ for different values of the anisotropy. Shown is the exact solution for the infinite system. The different panels are for $T/J=0.05$ (a), $0.5$ (b), and
$5$ (c).
\label{fig1}}
\end{figure}

Next, we discuss the effects of exchange anisotropy on the cooling rate.
Fig.~\ref{fig1} shows results for three selected values $\Delta=0.5$, $1$,
and $2$. Recall from Fig.~\ref{figXXZ} that,
at $\Delta = 0.5$, one crosses just one quantum phase transition
at $H_{c2}=1.5\,J$, the curves for $\Delta=1$ in Fig.~\ref{fig1}
start from the special
Heisenberg point at $H=0$ and cross the transition to saturation at
$H_{c2}=2\,J$, and the curves at $\Delta=2$ cut two quantum phase
transitions, namely $H_{c1} \approx 0.39\,J$ and $H_{c2} = 3\,J$.
For $T=0.05\,J$,
the quantum phase transitions at $H_{c2}$ and for $\Delta=2$ at
$H_{c1}$ are signaled by sign changes of the cooling rate $\Gamma_H$
from negative to positive values upon increasing field, see
Fig.~\ref{fig1}(a). Note that these zeros shift away from
the position of the zero-temperature phase transitions with
increasing temperature, as is evident in particular if one
takes into account the additional data for $\Delta=1$ shown
in Fig.~\ref{fig3}. Finally, there is a small structure at low
magnetic fields in the $\Delta=1$ curves of Figs.~\ref{fig3}(c) and
\ref{fig1}(a) which reflects the singular nature of the
Heisenberg point $(\Delta=1,H=0)$.

At the isotropic point of the antiferromagnetic $XXZ$ chain, the quantity 
$\Gamma_H/H$ shows singular behavior for $T\to 0$ if $H<T$. In fact, in the 
limit of $H\to 0$ the ratio reduces to the temperature derivative of the 
zero-field susceptibility $\chi(T)$
\be
\lim_{H\to 0}\frac{\Gamma_H(T)}H= -\frac{1}{C_H}\left( \frac{\partial \chi}
{\partial T}\right)_H \, , \label{ncr2}
\ee
which is known to show singular behavior. For $\Delta=1$ this function 
diverges like $\mbox{const.}\times [T \ln(T_0/T)]^{-2}$ where $T_0$ is a 
(non-universal) constant.\cite{Lukyanov} For $\Delta<1$ the function (\ref{ncr2}) diverges 
like $\mbox{const.}\times T^{4K-6}$ with Tomonaga-Luttinger parameter 
$K=\pi/(\pi-\gamma)>1$.\cite{Sirker}
For $\Delta>1$ a divergent behavior like $\mbox{const.}/T$ is observed.
The strongest divergence is hence exhibited for $\Delta=1$. For finite 
magnetic field $H$ we see a non-divergent behavior for $T\rightarrow 0$.

With increasing temperature, all features become broader.
Note that the temperature $T=0.5\,J$ shown in Fig.~\ref{fig1}(b)
is higher than the value of $H_{c1}$ at $\Delta = 2$
which explains why the sign change
around $H_{c1}$ in the $\Delta = 2$ curve disappears for
$T=0.5\,J$. For the even higher temperature $T=5\,J$ shown
in Fig.~\ref{fig1}(c), the cooling rate $\Gamma_H$ is positive
for all $H > 0$ (the zero at $H=0$ is enforced by the symmetry
under $H \to -H$). Note that $T=5\,J$ is bigger than $H_{c2}$
itself for all cases shown in Fig.~\ref{fig1}(c) which explains
why all features are washed out at this temperature.


\subsection{The case of infinitely large anisotropy:
            cooling rate for the Ising chain}

The case of infinitely large anisotropy
($\Delta \to \infty$ at $J_I=J \Delta =\mbox{const.}$) corresponds
to the Ising model with Hamiltonian
\bea
\mathcal{H}=J\sum _{i=1}^N s_i s_{i+1}-H \sum_{i=1}^Ns_i \, ,
\eea
where the classical variables $s_i$ take values  $\pm 1/2$ and for simplicity reasons we drop the subscript of the coupling $J_I$ in this section.
It is well known that the Ising chain can be solved exactly and completely analytically by the
transfer-matrix method (see for example Refs.\ \onlinecite{Huang,bax})
and even $\left(\partial M /\partial T\right)_H$ was investigated
recently.\cite{Sznajd08} However, we are not aware of explicit results
for the entropy $S$ and normalized cooling rate $\Gamma_H$ of the Ising
chain and therefore present them here.

We start from the free energy per lattice site of the Ising chain:
\begin{widetext}
\bea
f=-\frac{1}{\beta}\ln\left(e^{-\beta J/4} \left(\cosh (\beta H/2)+\sqrt{\sinh^2(\beta H/2)+e^{\beta J}} \right)  \right)
\, .
\eea
{}From this expression one can easily obtain simple analytic expressions
for all thermodynamic functions of the system. In particular one obtains
for the entropy per spin and the normalized cooling rate:
\bea
S&=&-\left(\frac{\partial f}{\partial T} \right)_H=\ln\left(e^{-\beta J/4}\left(c+Q
\right) \right)-\frac{-J/4 (c+Q)+s H/2  +\frac{s H/2  +J e^{\beta J}}{Q}}{T (c+Q)} \,
, \label{Si} \\
\Gamma_H &=& \frac{1}{T}\left(\frac{\partial T}{\partial H}
\right)_S=-\frac{\frac{\partial^2 f}{\partial T \partial H}}{T\frac{\partial^2
f}{\partial T^2 }}=\frac{ 1/2(c H/2 - s J/2 )(c+Q)^2}{1/2 J^2 s^2 Q+1/4 J^2 c (s^2+Q^2)+H/2 (c H/2
-J s)(c+Q)^2} \, , \label{Gi}
\eea
where
\begin{equation}
c=\cosh (\beta H/2)\,, \qquad 
s=\sinh (\beta H/2)\,, \qquad 
Q=\sqrt{\sinh^2 (\beta H/2)+e^{\beta J}} \, . \label{not}
\end{equation}
\end{widetext}
The $T=0$ limit of the entropy (\ref{Si}) is generically $S=0$, except
for $H = \pm J$, where one finds $S=\ln\left(\frac{1+\sqrt{5}}{2}\right)
= 0.4812\ldots$ (see also Ref.~\onlinecite{MeYa78}).
This reflects the macroscopic ground-state
degeneracy of the Ising model at the saturation field $H_c=J$.\cite{Bonner1962}
Remarkably, the above transfer-matrix solution is closely related to
the hard-dimer description of certain highly frustrated one-dimensional
quantum antiferromagnets.\cite{zhhon,Derzhko07,DeRi04,ZhiT04,ZhiT05}

\begin{figure}[tb!]
\includegraphics[width=\columnwidth]{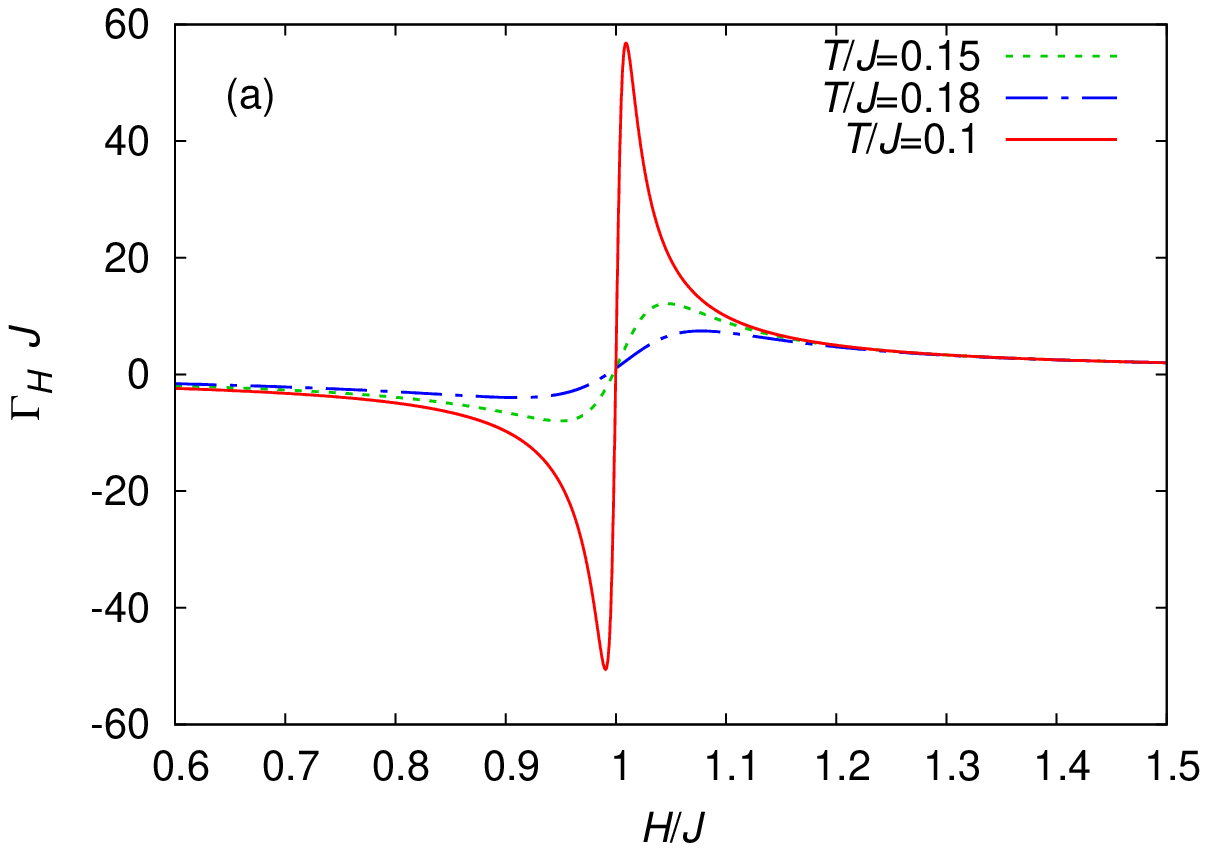}
\includegraphics[width=\columnwidth]{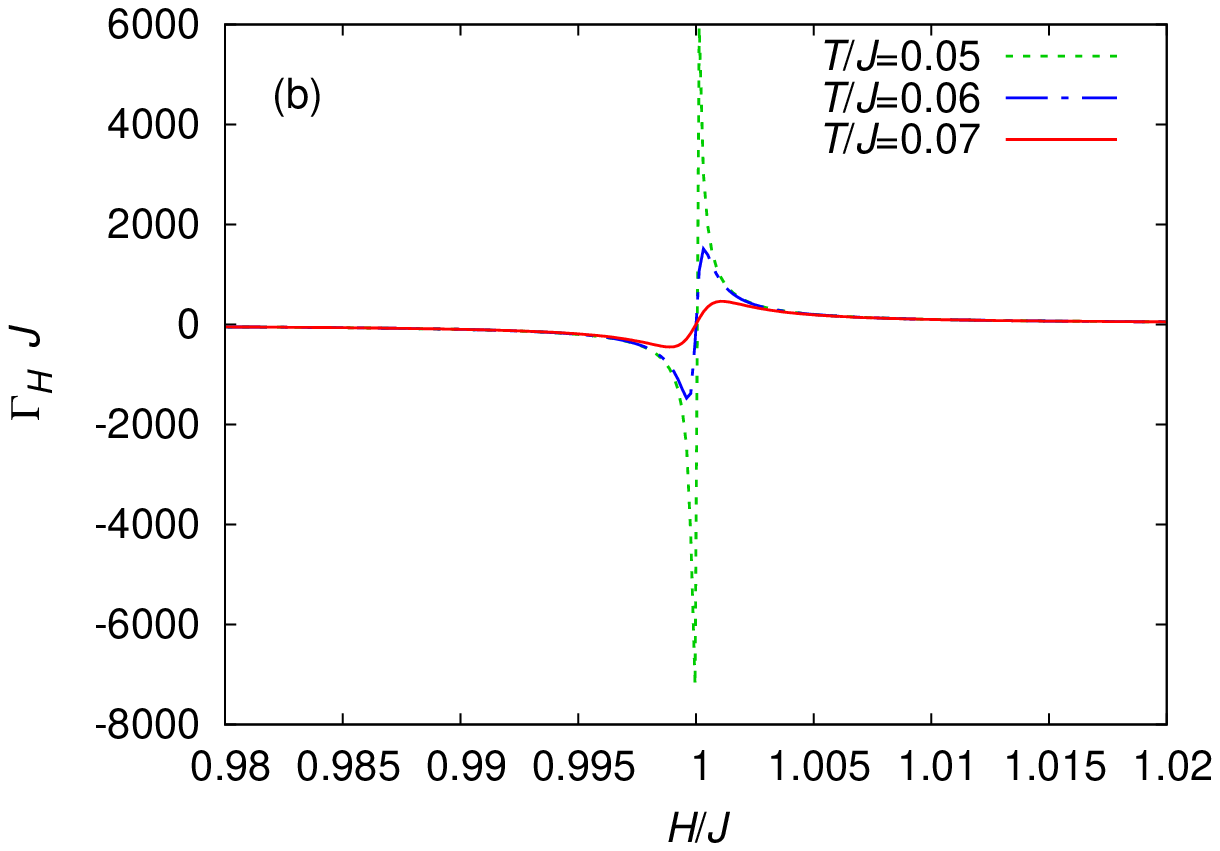}
\caption{(Color online) Cooling rate $\Gamma_H$ of the Ising model ($\Delta \to \infty$
limit of the $XXZ$ model). Shown is the exact solution for the infinite system. The different panels show the cooling rate for a wider range of temperatures (a) and the behavior at low temperatures in the vicinity of  the critical point (b).
\label{fig2}}
\end{figure}

Fig.\ \ref{fig2} shows the cooling rate dependence
on the magnetic field for the Ising chain obtained from the
exact solution. The main difference from the $XXZ$ case are
the very sharp and pronounced positive and negative peaks
at the critical value of the magnetic field. The magnitude of
these peaks grows rapidly with decreasing temperature.
This behavior is a direct consequence of the
anomalous zero-temperature entropy of the Ising chain at $H=J$.\cite{Bonner1962}

A bit further away from the QCP at $H_c=J$, we can make contact
with the argumentation of Ref.~\onlinecite{Garst}.
The Gr\"{u}neisen ratio, which in our case is the cooling rate $\Gamma_H$, shows 
divergent behavior close to the QCP at $H_c$ and changes sign when
the magnetic field crosses it.
For $H \ne H_c$, the divergent behavior obeys the universal scaling law\cite{Zhu,Garst}
\begin{eqnarray}
\Gamma_H\left(T\rightarrow 0, H\right)=-G_H\frac{1}{H-H_c}, \label{scaling}
\end{eqnarray}
where $G_H$ is a universal amplitude. Detailed analysis of the Ising case yields
the exact analytic form of the cooling rate
at extremely low temperatures,
$\Gamma_H^{\rm Ising} \left( T \rightarrow 0\right)=\frac{1}{H-J}$
for $T \ll \abs{H-J}$, {\it i.e.}, the value $G_H=-1$ expected
for a $\mathbb{Z}_2$-symmetry in one dimension.\cite{Garst,zhhon}

\subsection{Entropy}

\begin{figure}[tb!]
\includegraphics[width=\columnwidth]{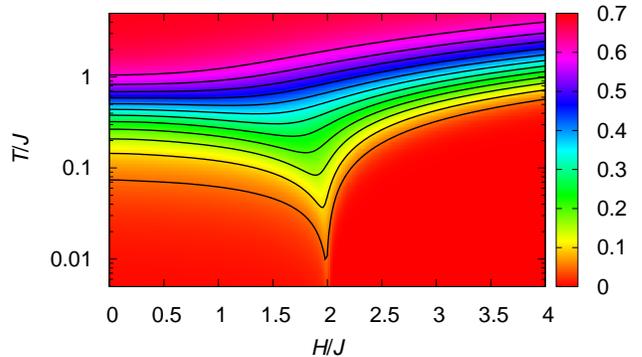}
\caption{(Color online) Entropy per spin $S(H,T)$ for $\Delta=1$.
The isentropes are for $S=0.05,0.1,\ldots,0.6$.
\label{fig4}}
\end{figure}

\begin{figure}[tb!]
\includegraphics[width=\columnwidth]{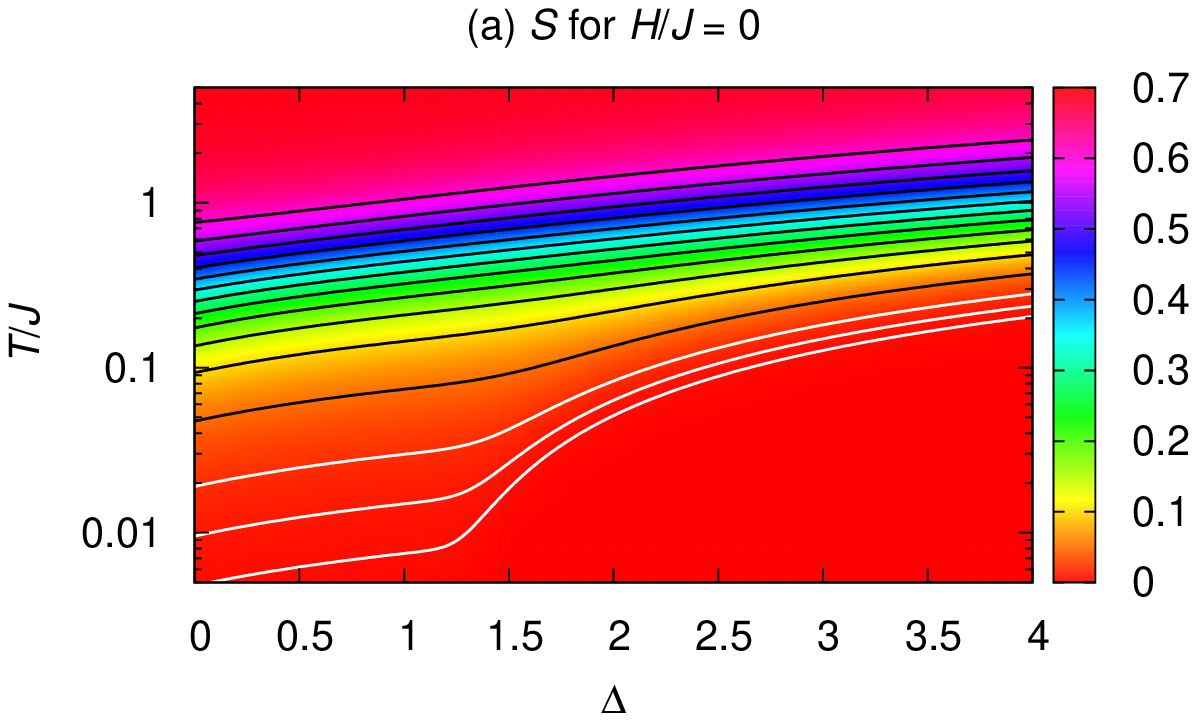}
\includegraphics[width=\columnwidth]{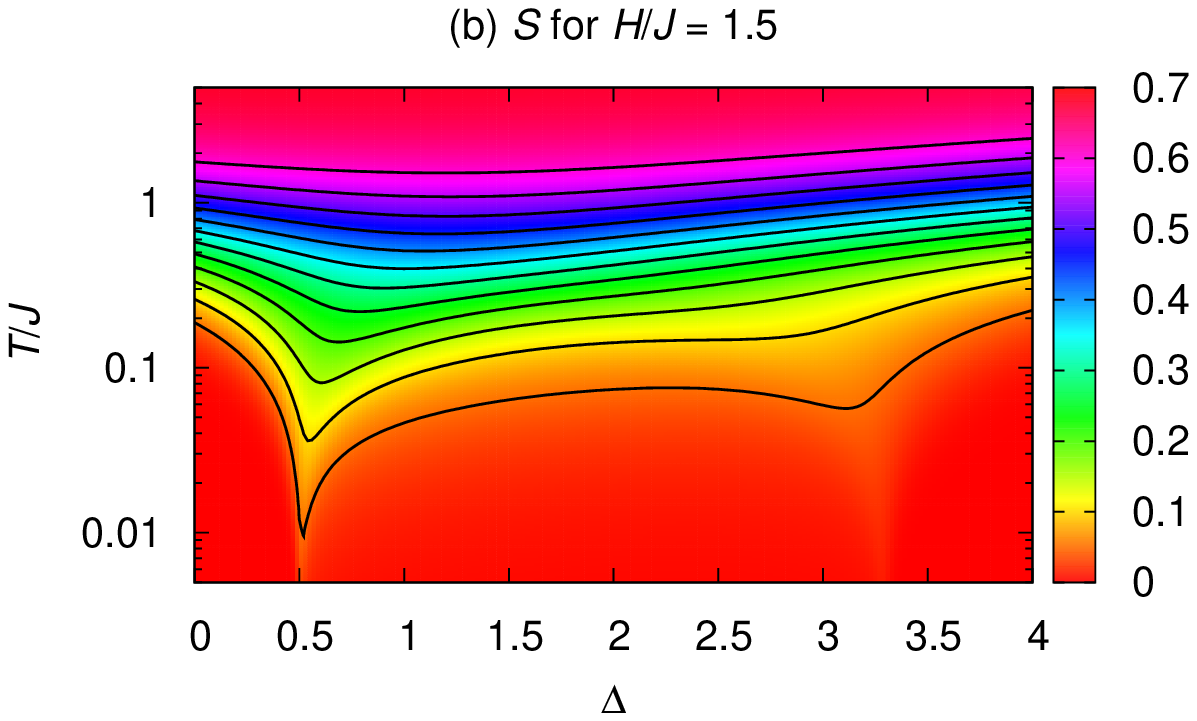}
\includegraphics[width=\columnwidth]{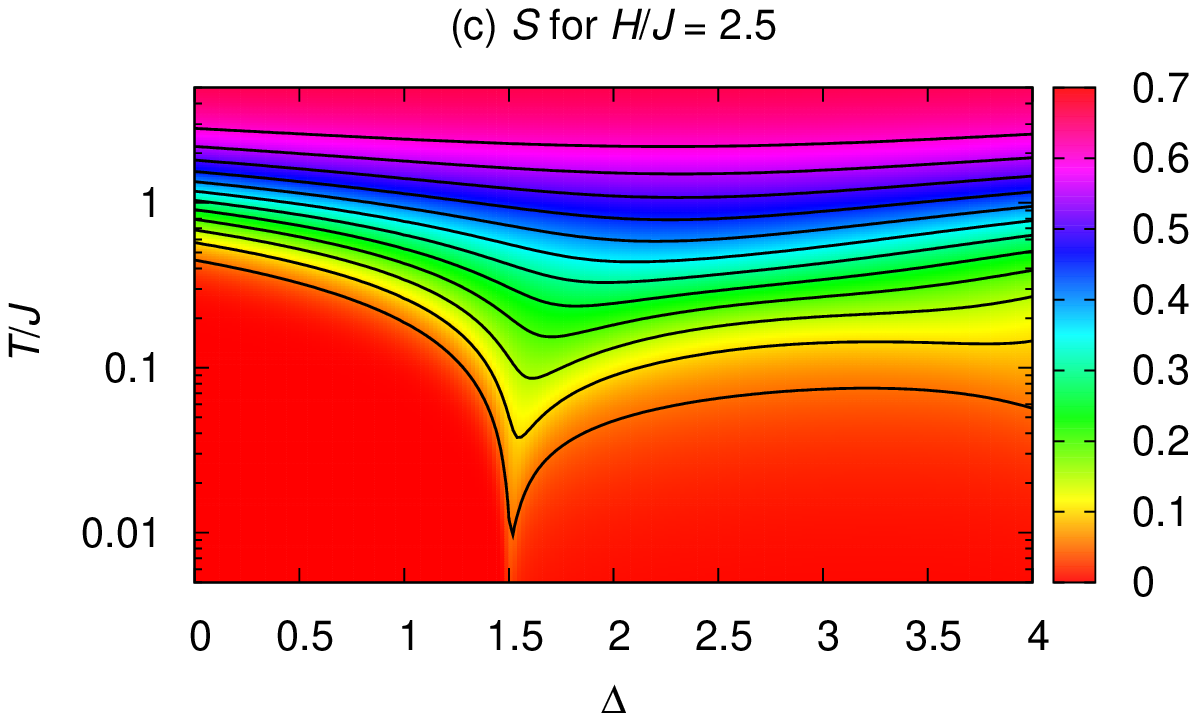}
\caption{(Color online) Entropy per spin $S(\Delta,T)$
for different values of $H=0$ (a), $1.5\,J$ (b), and $2.5\,J$ (c).
The black lines are the isentropes for $S=0.05,0.1,\ldots,0.6$.
Panel (a) contains three additional isentropes with $S=0.005$, $0.01$,
and $0.02$ which are shown by white lines.
\label{fig5}}
\end{figure}

Finally, we take a look at the entropy $S$.
As far as we are aware, there has been only one previous attempt
of an exact computation\cite{Bonner1977}
of the entropy and accordingly the isentropes of the $XXZ$ chain
in a magnetic field. Inspection of the critical fields
$H_{c1}$ and $H_{c2}$ indicates that
the result of Ref.~\onlinecite{Bonner1977} corresponds to an anisotropy
$\Delta \approx 4.75$. The main technical difference between
Ref.~\onlinecite{Bonner1977} and the present work is that
Ref.~\onlinecite{Bonner1977} worked with an infinite hierarchy of
equations for $\Delta > 1$ which need to be truncated while we have
a closed finite set of equations for all values of $\Delta$.
In the case $|\Delta|<1$
the structure of the NLIE \eqref{uu} is independent of $\Delta$
in contrast to the usual TBA equations, which allows to easily calculate
the entropy as a function of the anisotropy.

Fig.~\ref{fig4} shows the entropy and the isentropes for the
spin-1/2 Heisenberg chain in the $H$-$T$ plane. This result
agrees with previous ED for $N=20$ sites.\cite{zhhon} However,
the ED results of Ref.~\onlinecite{zhhon} suffered from finite-size
effects, in particular for low temperatures and $H < H_{c2} = 2\,J$.
By contrast, Fig.~\ref{fig4} shows results for the thermodynamic
limit. Fig.~\ref{fig5} shows a similar plot of the entropy and
isentropes with varying anisotropy $\Delta$ but now at a fixed
magnetic field $H$. The quantum phase transitions
at $H_{c1}$ and $H_{c2}$ are reflected in Figs.~\ref{fig4} and
\ref{fig5} by minima of the isentropes as a function of $H$ or
$\Delta$, or equivalently
maxima of the entropy at a low but constant temperature.
The only exception is Fig.~\ref{fig5}(a) where one observes no such
clear signature of the Kosterlitz-Thouless transition at
$\Delta=1$ for $H=0$. Before we discuss this case in more detail,
it is useful to examine the low-temperature asymptotics of the
entropy $S(T)$ at otherwise fixed parameters $H$ and $\Delta$.

Between the two quantum critical points, that is for
$H_{c1} < H < H_{c2}$, the low-energy theory is a Luttinger
liquid (compare remarks at the beginning of Sec.~\ref{sec:Heis}).
It is well known that the specific heat $C$ of a Luttinger
liquid is linear in $T$.\cite{Giamarchi} Due to the
relation $C=T\,(\partial\,S/\partial T)$  (see, e.g., Ref.~\onlinecite{rev2})
and because of $S(T=0) = 0$, the entropy
of a Luttinger liquid is identical to its specific heat
and in particular also linear
\begin{equation}
S = \frac{\pi}{3\,v}\,T \quad \mbox{for } H_{c1} < H < H_{c2} \, ,
\label{eq:STlin}
\end{equation}
where $v$ is the velocity of the excitations.

The cases $H=H_{c1}$ and $H_{c2}$ are instances of a quantum phase
transition in one dimension which preserves a $U(1)$-symmetry.
In this case, the universal low-temperature asymptotics is predicted to
follow a square root\cite{Bonner1972,zhhon,Zhu,Garst}
\begin{equation}
S \propto \sqrt{\frac{T}{J}} \quad  \mbox{for } H=H_{c1} \mbox{ or } H_{c2} \, .
\label{eq:STsqrt}
\end{equation}
Finally, in the gapped cases $0 \le H < H_{c1}$ or $H > H_{c2}$, we expect
activated behavior
\begin{equation}
S \propto \exp(-G/T) \, ,
\label{eq:STgap}
\end{equation}
where $G$ is the gap in the excitation spectrum. Closer inspection
of the data shown in Figs.~\ref{fig4} and \ref{fig5}
indeed verifies Eqs.\ (\ref{eq:STlin}), (\ref{eq:STsqrt}),
or (\ref{eq:STgap}), respectively. Note that while
the asymptotic behavior is $S \to 0$ for $T \to 0$ in
all three cases, the decay is slowest exactly at
the quantum phase transition, see Eq.~(\ref{eq:STsqrt}).
This naturally yields a maximum of the entropy $S$
if the quantum phase transition is crossed by varying
the parameters $H$ or $\Delta$ at a fixed temperature $T > 0$.

\begin{figure}[tb!]
\includegraphics[width=\columnwidth]{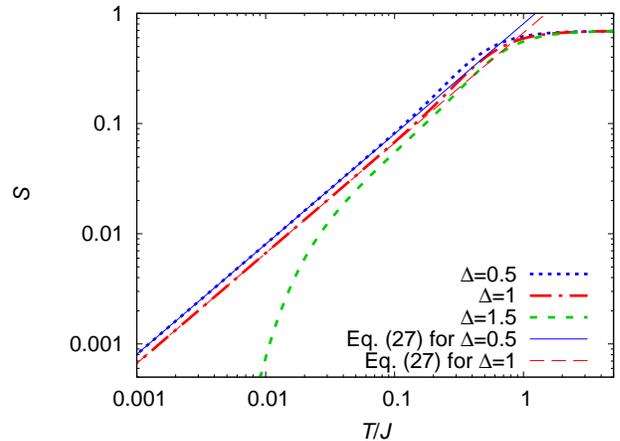}
\caption{(Color online) Entropy per spin $S$ of the
$XXZ$ chain at $H=0$ for $\Delta=0.5$, $1$, and $1.5$.
For $\Delta=0.5$ and $1$ we also show
the Luttinger-liquid expression Eq.~(\ref{eq:STlin})
with the velocity $v$ given by Eq.~(\ref{eq:ValV}).
\label{fig7}}
\end{figure}

The case $\Delta=1$, $H=0$ is an exception to this
general scenario. To first approximation, this
point behaves like a Luttinger liquid. The fact that one is at a 
quantum critical point with marginally irrelevant
operators gives rise to higher-order logarithmic corrections 
in the free energy and specific heat.\cite{klu98,klu00}
Consequently, one also expects just higher-order
logarithmic corrections to the entropy. To test this scenario,
we can use the fact that at $H=0$ the velocity $v$
which enters (\ref{eq:STlin}) is known exactly
(see, e.g., Refs.~\onlinecite{Haldane,Giamarchi}):
\begin{equation}
v = \frac{\pi\,\sin{\gamma}}{2\,\gamma}\,J
\label{eq:ValV}
\end{equation}
with $\Delta = \cos\gamma$ as above.
Fig.~\ref{fig7} shows that insertion of these values of $v$
into the Luttinger-liquid expression Eq.~(\ref{eq:STlin})
yields indeed the correct low-temperature asymptotics
of the entropy per spin at $H=0$ not only for $\Delta=0.5$
but also for $\Delta=1$. In fact, the
higher-order logarithmic corrections to
the Luttinger liquid asymptotics (\ref{eq:STlin})
which are expected at $\Delta=1$ and $H=0$ are so small
that they have no visible effect.

A gap $G$ opens for $\Delta > 1$ at $H=0$,
but because the phase transition is a Kosterlitz-Thouless
transition,\cite{KT73,Kosterlitz} this gap is exponentially small close
to $\Delta=1$.\cite{CG66,yang_yang_1966_3}
Accordingly, close to $\Delta=1$
one has to go to very low temperatures in
order to observe the crossover from
(\ref{eq:STlin}) to (\ref{eq:STgap}).
This is illustrated by the $\Delta = 1.5$ curve
in Fig.~\ref{fig7}. In the concrete case $\Delta = 1.5$,
the value of the gap is $G\approx 0.087\,J$.
Accordingly, the exponential decay (\ref{eq:STgap})
can be observed only for temperatures $T \lesssim 0.05\,J$
while at higher temperatures the behavior of $S(T)$ remains
approximately linear.

The combined effect of all these observations
is that just a small kink develops in the
low-temperature isentropes of Fig.~\ref{fig5}(a)
whose position shifts very slowly to the quantum
critical point $\Delta=1$ for $T \to 0$.

\section{Conclusion and Perspectives}


Motivated by recent measurements of the magnetic cooling rate
in a spin-1/2 Heisenberg chain compound,\cite{Tsui}
we have presented an exact computation of the entropy
and the magnetic cooling rate of the antiferromagnetic
spin-1/2 $XXZ$ chain in the thermodynamic limit $N \to \infty$.
Furthermore, we have performed complementary numerical computations
for the cooling rate of finite Heisenberg chains, namely exact
diagonalization (ED) for small systems and QMC simulations for somewhat longer
chains. We have demonstrated that we can obtain excellent approximations
to the exact result with a combination of both numerical methods.
On the one hand, this serves as a consistency check of our computations.
On the other hand, we are now in a position to apply a combination
of ED and QMC to some sign-free situations, like the spin-1/2 Heisenberg
model on the square and simple cubic lattices where the exact methods
are no longer applicable.

We have used the exact result for the entropy to illustrate that
field-induced quantum phase transitions give rise to
maxima of the low-temperature entropy, or equivalently
minima of the isentropes. This leads to cooling during
adiabatic (de)magnetization processes where the lowest
temperature is reached close to the quantum phase transition.
As a consequence, we find a zero for the magnetic
cooling rate at the phase transition and large
positive (negative) values of the normalized cooling
rate (\ref{ncr}) for magnetic fields slightly
above (below) the critical field.

The low-temperature asymptotics of the entropy $S$ is exponentially activated
in the gapped phases, linear in $T$ in the gapless Luttinger liquid
regions, and follows the square-root behavior
(\ref{eq:STsqrt}) at the field-induced quantum phase transitions.
These asymptotic forms of $S(T)$ are expected to be universal for field-induced
phase transitions in one-dimensional systems
with $U(1)$-symmetry,\cite{Bonner1972,zhhon,Zhu,Garst}
but can be particularly clearly verified with the aid of an
exact solution.

The general features of the entropy should not depend on
the specific choice of the magnetic field $H$ as control
parameter and indeed similar behavior is found as a
function of the exchange anisotropy $\Delta$. An exception
is just the quantum phase transition at $H=0$, $\Delta=1$
with $\Delta$ as a control parameter. Because this is
a Kosterlitz-Thouless transition, only weak signatures are
observed in the finite-temperature entropy.

Finally, closed-form expressions were derived in the Ising limit using the
transfer matrix method.\cite{Huang,bax} We have observed remarkably
large magnetic cooling rates close to the field-induced critical point
of the Ising chain.
In fact, the transfer-matrix solution is closely related to a
low-energy description of highly frustrated one-dimensional
quantum antiferromagnets,\cite{zhhon,Derzhko07,DeRi04,ZhiT04,ZhiT05}
where enhanced cooling rates are observed as well.

\acknowledgments

V.O.\ would like
to thank the Institut f\"{u}r Theo\-re\-ti\-sche Physik, Universit\"{a}t G\"{o}ttingen
and Universit\"{a}t Wuppertal for hospitality during the course of this
work. This research stay was supported by the German Science Foundation
(DFG). Furthermore, A.H.\ is grateful to the DFG for financial
support by a Heisenberg fellowship (grant no.\ HO~2325/4-1).
V.O.\ was also supported by the grants UCEP-CRDF-06/07 and ANSEF-1518-PS.
C.T.\ would like to acknowledge support by the research program of the Graduiertenkolleg 1052 funded by the DFG.

\end{document}